\documentstyle[twoside,fleqn,espcrc2]{article}

\input psfig.tex
\def\mycite#1{$\,$\cite{#1}}

\def\be{\begin{equation}}
\def\ee{\end{equation}}
\def\bea{\begin{eqnarray}}
\def\eea{\end{eqnarray}}

\def\ma{m_{\rm a}}
\def\fa{f_{\rm a}}

\def\dalemb#1#2{{\vbox{\hrule height .#2pt
\hbox{\vrule width.#2pt height#1pt \kern#1pt\vrule width.#2pt}
\hrule height.#2pt}}}

\def\tdot{\kern -8.5pt{}^{{}^{\hbox{...}}}}
\def\dotprime{\kern -8.0pt{}^{{}^{\hbox{.}~\prime}}}

\def\lapp{\hbox{$ {     \lower.40ex\hbox{$<$}
                   \atop \raise.20ex\hbox{$\sim$}
                   }     $}  }
\def\gapp{\hbox{$ {     \lower.40ex\hbox{$>$}
                   \atop \raise.20ex\hbox{$\sim$}
                   }     $}  }

\def\marbul{\strut\vadjust{\kern-2pt$\bullet$}}

\mathchardef\less="321C

\def\smallskip{\vskip 10pt}
\def\microeV{\mu\hbox{\rm eV}}

\title{On the Origin of Dark Matter Axions}
\author{E.P.S.~Shellard and R.A.~Battye
\address{Department of Applied Mathematics and Theoretical Physics, University of Cambridge, \\ Silver Street, Cambridge, CB3 9EW, U.K.}}

\begin{document}

\begin{abstract}

We discuss the possible sources of dark matter axions in the early universe.  
In the standard thermal scenario, an axion string network forms at 
the Peccei-Quinn phase transition $T\sim \fa$ and then radiatively decays
into a cosmological background of axions; to be the dark matter, these axions 
must have a mass $\ma \sim 100\,\mu$eV with specified large uncertainties.
An inflationary phase with a reheat temperature below the PQ-scale 
$T _{\rm reh} \lapp \fa$ can also produce axion strings through quantum 
fluctuations, provided that the 
Hubble parameter during inflation is large $H_1 \gapp \fa$;
this case again implies a dark matter axion mass $\ma \sim 100\,\mu$eV.  For a
smaller Hubble parameter during inflation $H_1 \lapp \fa$, `anthropic tuning'
allows dark matter axions to have any mass in 
a huge range below $\ma\lapp 1\,$meV.

\end{abstract}

\maketitle
  
\section{INTRODUCTION}

The axion has remained a popular dark matter candidate because of its
enduring motivation as an elegant solution to the strong CP-problem\mycite{PecQui77}.  Despite early hopes of discovery, 
it turned out that in order to be consistent with accelerator
searches and astrophysics, the axion must be nearly `invisible' and extremely 
light. Its couplings and mass are inversely proportional to the (large)
Peccei-Quinn scale $\fa$ as in
\be
\ma = 6.2 \times 10^{-6}\hbox{eV} \left ( 10^{12}\hbox{GeV}\over \fa\right)\,.
\label{axionmass}
\ee
Accelerator constraints have been largely superseded by those from 
astrophysics; because the axion is so weakly coupled, volume effects can 
compete with other surface and convective stellar energy loss mechanisms.
The strongest astrophysical constraints on 
the axion mass derive from studies of supernova 1987a and conservative
estimates yield
$\ma ~\lapp~ 10\,\hbox{meV}$\mycite{Raf97}.  The present programme of 
large-scale axion search experiments\mycite{Hagetal98} are
sensitive to a  mass range $m_a \sim 1\hbox{--}10\,\microeV$, which has been 
chosen for a variety of historical and technological reasons. 
Our primary focus here, however, is not on constraints on 
the {\it viable} axion mass range, but rather on efforts
to predict the mass of a {\it dark matter} axion from cosmology.

\section{STANDARD AXION COSMOLOGY}

\noindent The cosmology of the axion is determined by the two 
energy scales $\fa$ and $\Lambda_{\rm QCD}$.  The first important event
is the Peccei-Quinn phase transition which is broken at a high temperature
$T\sim \fa \gapp 10^9$GeV.  This creates the axion,
at this stage an effectively massless pseudo-Goldstone boson, 
as well as a network of axion 
strings\mycite{VilEve82} 
which decays gradually into a background cosmic axions\mycite{Dav86}.
(Note that one can engineer models in which an 
inflationary epoch interferes with the effects of 
the Peccei-Quinn phase transition, as we shall discuss in the next section.)  
At a much lower temperature $T\sim \Lambda_{\rm QCD}$
after axion and string formation, instanton effects `switch on', the axions
acquire a small mass, domain walls form\mycite{Sik82} between the 
strings\mycite{VilEve82} 
and the complex
hybrid network annihilates into axions in about one Hubble time\mycite{She86}.  

There are three possible mechanisms by which axions are produced in the
`standard thermal scenario':  (i) thermal production, (ii) axion string
radiation and (iii) hybrid defect annihilation when $T=\Lambda_{QCD}$.  
Axions consistent with the astrophysical bounds must decouple from 
thermal equilibrium very early;
their subsequent 
history and number density is analogous to the decoupled neutrino,
except that unlike a 100eV massive neutrino, thermal axions cannot hope 
to dominate the universe with $\ma\lapp 10$meV.  
We now turn to the two dominant axion production mechanisms, but first we 
address an important historical digression.

\subsection{Misalignment misconceptions} 
The original papers on axions suggested 
that axion production
primarily occurred, not through the above mechanisms, but 
instead by `misalignment' effects at the QCD phase transition
\mycite{PAD83}.  Before the
axion mass `switches on', the axion field $\theta$ takes random 
values throughout
space in the range 0 to $2\pi$; it is the phase of the PQ-field lying at the 
bottom of a $U(1)$ `Mexican hat' 
potential.  However, afterwards the potential becomes tilted and 
the true minimum 
becomes $\theta=0$, so the field in the `misalignment' picture begins to 
coherently oscillate about this minimum; this homogeneous mode corresponds
to the `creation' of zero momentum axions. Given an initial
rms value $\theta_{\rm i}$ for these oscillations, it is 
relatively straightforward to 
estimate the total energy density in zero momentum axions and compare 
these to the present mass density of the universe (assuming
a flat $\Omega=1$ FRW cosmology)\mycite{PAD83,Tur86}:
\be
\Omega_{{\rm a},{\rm hom}}  ~\approx ~
 2 \,\Delta h^{-2}\;\theta_{\rm i}^2 
{\rm f}(\theta_{\rm i})\,
\bigg{(}{10^{-6}{\rm eV}\over \ma}\bigg{)}^{1.18}  
\label{homcount}
\ee
where $\Delta\approx 3^{\pm 1}$ accounts for both model-dependent axion 
uncertainties and those due to the nature of the QCD phase 
transition, and $h$ is the rescaled Hubble parameter. 
The function ${\rm f} (\theta)$ is an anharmonic correction for 
fields near the top of the potential close to  unstable 
equilibrium $\theta\approx\pi$, that is, 
with ${\rm f}(0)=1$ at the base $\theta\approx 0$ and diverging 
logarithmically 
for $\theta\rightarrow\pi$\mycite{SheBat98}.
If valid, the estimate (\ref{homcount}) would imply a 
constraint $\ma\gapp 5\mu {\rm eV}$ for the anticipated
thermal initial conditions with $\theta_{\rm i} = 
{\cal O}(1)$\mycite{PAD83,Tur86}. 

As applied to the thermal scenario, the expression (\ref{homcount}) 
is actually a very considerable
underestimate for at least two reasons:  First, the axions are not `created' by 
the mass `switch on' at $t=t_{QCD}$, they are already there 
with a specific
momentum spectrum $g(k)$ determined by dynamical mechanisms prior to this time.
The actual axion number obtained from $g( k)$ is much 
larger than the rms average assumed in (\ref{homcount}) 
which ignores the true particle content.  Secondly,
this estimate was derived before much stronger topological 
effects were realized, notably the presence of 
axion strings and domain walls. In any case, these
nonlinear effects complicate the oscillatory behaviour considerably,
implying that the homogeneous estimate (\ref{homcount}) is poorly motivated.

\subsection{Axion string network decay} Axions and axion strings are inextricably intertwined.  Like
ordinary superconductors or superfluid $^4$He, axion models contain 
a broken $U(1)$-symmetry and so there exist vortex-line solutions.  Combine
this fact with the Peccei-Quinn phase transition, which means the field 
is uncorrelated beyond the horizon, and a random network of axion
strings must inevitably form.  An axion string corresponds to a non-trivial
winding from $0$ to $2\pi$ of the axion field $\theta$ around the bottom of its 
`Mexican hat' potential.  It is a global string with 
long-range fields, so its energy per unit length $\mu$ has a logarithmic 
divergence which is cut-off by the string curvature radius $R\lapp t$,
that is, $\mu \approx 2\pi \fa^2 \ln (t/\delta)\,,$
where the string core width is $\delta \approx \fa^{-1}$.  The axion string,
despite this logarithmic divergence, 
is a strongly localized object; if we have a string
stretching across the horizon at the QCD 
temperature, then $\ln(t/\delta)\sim 65$ and over 95\% of its 
energy lies within a tight cylinder enclosing only 0.1\% of 
the horizon volume.  To first order, then, the string behaves like 
a local cosmic string, a fact that can be established by a precise
analytic derivation and careful comparison with numerical 
simulations\mycite{BatShe94a}.  

After formation and a short period of damped evolution, the axion string
network will evolve towards a scale-invariant regime with a fixed number of
strings crossing each horizon volume (for a cosmic string 
review see ref.\mycite{VilShe94}). 
This gradual demise of the network
is achieved by the production of small loops which oscillate relativistically 
and radiate primarily into axions.
The overall density of strings splits neatly into
two distinct parts, long strings with length $\ell > t$ and a population of 
small loops $\ell < t$, that is, $\rho=\rho_{\infty}+\rho_{L}$.
High resolution numerical simulations confirm 
this picture of string 
evolution and suggest that the long 
string density during the radiation era is $\rho_\infty \approx 
13\mu/t^2$\mycite{BBAS90}. 
To date, analytic descriptions of the loop distribution have used the 
well-known string `one scale' model, which predicts a
number density of loops defined as $\mu\ell\,n(\ell,t)\,d\ell=\rho_{L}
(\ell,t)d\ell$ in the interval $\ell$ 
to $\ell+d\ell$ to be given by 
\be
n(\ell,t)={4\alpha^{1/2}(1+\kappa/\alpha)^{3/2}
\over (\ell+\kappa t)^{5/2} t^{3/2}}
\,,\label{numdenloop}
\ee 
where $\alpha$ is the typical loop creation size relative to the horizon 
and $\kappa \approx 65/[2\pi \ln (t/\delta)]$ is the loop radiation 
rate\mycite{BatShe94b}. Once formed at $t=t_0$ with length $\ell_0$, a typical loop 
shrinks linearly as it decays into axions $\ell = \ell_0 - \kappa(t -t_0)$.
The key uncertainty in this treatment is the loop creation size $\alpha$, but 
compelling heuristic arguments place it near the radiative backreaction scale, 
$\alpha \sim \kappa$.  (If this is the case, we note that the loop contribution 
is over an order of magnitude larger than direct axion radiation from long
strings.)

String loops oscillate with a period $T=\ell/2$ and radiate
into harmonics of this frequency (labelled by $n$), just like other 
classical sources.  Unless a loop has a particularly 
degenerate trajectory, it will have a radiation spectrum $P_n \propto n^{-q}$
with a spectral index $q>4/3$, that is, the spectrum is dominated by 
the lowest available modes.\footnote{Historically, there has been some debate on 
the radiation spectrum issue, but the reader is referred elsewhere for
further details\mycite{BatShe98b}.} Given the loop density (\ref{numdenloop}), we can 
then calculate the spectral number density 
of axions $dn_{\rm a}/d\omega$,
which turns out to be 
essentially independent of the exact loop radiation spectrum 
for $q>4/3$.   From this expression we can integrate over $\omega$
to find the total axion number at the time $t_{\rm QCD}$, that is, when the axion 
mass `switches on' and the string network annihilates. Subsequently,
the axion number will be conserved, so we can find the number-to-entropy ratio
and project forward to the present day.  Multiplying the present number
density by the axion 
mass $\ma$ yields the overall axion string contribution to the density of 
the universe\mycite{BatShe94b}:
\be
\Omega_{\rm a,string}\approx 110\Delta h^{-2}
\bigg{(}{10^{-6}\hbox{eV}\over \ma}\bigg{)}^{1.18}f(\alpha/\kappa)
\,,\label{stringbound}
\ee
where
\be
f(\alpha/\kappa)=\bigg{[}\bigg{(}1+{\alpha\over\kappa}\bigg{)}^{3/2}-1\bigg{]}\,.
\ee
The key additional 
uncertainty from the string model is the ratio $\alpha / \kappa\sim {\cal O}(1)$,
which  
should be clearly distinguished from 
particle physics and cosmological uncertainties inherent in $\Delta$ and $h$
(which appear in all estimates of $\Omega_{\rm a}$).  With a Hubble parameter 
near $h=0.5$,
the string estimate (\ref{stringbound}) tends to favour a dark matter axion with a mass 
$\ma \sim 100\mu$eV, as we shall discuss in the conclusion.
A comparison with (\ref{homcount}) confirms that
$\Omega_{\rm a,string}$ is well over an order of magnitude 
larger than the `misalignment' contribution.

\subsection{Hybrid defect annihilation}
Near the QCD phase transition the axion acquires a mass and network
evolution alters dramatically because domain walls form. 
Large field variations around the strings collapse into 
these domain walls, which subsequently begin to dominate over 
the string dynamics.  This occurs when the wall 
surface tension $\sigma$ becomes comparable to the 
string tension due to the typical curvature $\sigma\sim\mu/t$. 
The demise of the hybrid string--wall network proceeds rapidly, as
demonstrated numerically\mycite{She86}.  The strings frequently intersect and
intercommute with the walls, effectively `slicing up' the network into small
oscillating walls bounded by string loops.  Multiple self-intersections will reduce
these pieces in size until the strings dominate the dynamics again and decay
continues through axion emission.

An order-of-magnitude estimate of the demise of the string--domain wall
network indicates that there is an additional contribution\mycite{Lyt92} 
\be
\Omega_{\rm a,dw}\sim{\cal O}(10)\Delta h^{-2}
\bigg{(}{10^{-6}\hbox{eV}\over \ma}\bigg{)}^{1.18}\,.
\ee
This `domain
wall' contribution is ultimately due to loops which are created at 
the time $\sim
t_{\rm QCD}$. Although the resulting loop density will be similar 
to (\ref{numdenloop}),
there is not the same accumulation from early times, so it is likely to be
subdominant\mycite{BatShe94b} relative to (\ref{stringbound}). 
More recent work,\mycite{Nag97}
questions this picture by suggesting that the walls stretching between 
long strings dominate and will produce a contribution anywhere in 
 the wide range $\Omega_{\rm a,dw} 
\sim (1\hbox{--}44) \Omega_{\rm a,string}$; however, this assertion 
requires stronger quantitative support.
Overall, like most effects,\footnote{We note briefly that it is also 
possible to weaken any axion mass bound through 
catastrophic entropy production between the QCD-scale and nucleosynthesis,
that is, in the timescale range $10^{-4}s \lapp t_{\rm ent} \lapp 10^{-2}s$.  
Usually this involves the energy density of 
the universe becoming temporarily dominated by an exotic massive particle 
with a tuned decay timescale.} the domain wall
contribution will serve to further strengthen the string  
bound (\ref{stringbound}) on the axion.

Up to this point we have only considered the simplest axion models 
with a unique
vacuum $N=1$, so what happens when $N>1$?  In this case, any strings 
present become attached to $N$ domain walls at the QCD-scale.  Such a 
network `scales' rather than annihilates, and so it is cosmologically
disastrous being incompatible (at the very least) with CMB 
isotropy.

\section{INFLATIONARY AXION MODELS}

\def\Treh{T_{\rm reh}}

\noindent The relationship between inflation and dark matter axions is
rather mysterious.  Its significance depends on the magnitude of the Peccei-Quinn scale 
$\fa$ relative to two key inflationary parameters, (i) the reheat temperature
of the universe $T_{\rm reh}$ at the end of inflation and (ii) the Hubble parameter
$H_1$ as the observed universe first exits the horizon during inflation. 
Inflation is irrelevant to the axion if $\Treh\gapp\fa$ because, in this case, the 
PQ-symmetry is restored and the universe returns to the `standard thermal
scenario' in which axion strings form and the estimate (\ref{stringbound}) pertains.
Consider, then, the two inflationary axion scenarios with $T_{\rm reh} \lapp \fa$.

\subsection{Anthropic misalignment and quantum fluctuations \it (Case 1: 
$H_1<\fa$)}
In an inflationary model for which $\fa>H_1>T_{\rm reh}$, the 
$\theta$-parameter or axion angle will be set homogeneously 
over large inflationary 
domains before inflation finishes\mycite{Pi84}.  In this case, the 
whole observable
universe emerges from a single Hubble volume in which this parameter has some
fixed initial value $\theta_{\rm i}$. Because the
axion remains out of thermal equilibrium for large $\fa$, subsequent evolution 
and reheating does not disturb $\theta_{\rm i}$ until 
the axion mass `switches on' at  $T\sim\Lambda_{\rm QCD}$.  Afterwards,
the field begins
to oscillate coherently, because it is  
misaligned by the angle $\theta_{\rm i}$ from the true minimum $\theta=0$. 
This homogeneous mode corresponds to a background of zero momentum axions 
and it is the one  
circumstance under which the misalignment formula (\ref{homcount})
actually gives an accurate estimate of the relative axion density 
$\Omega_{\rm a}$.

By considering the dependence $\Omega_{\rm a,h} \propto \theta_{\rm i}^2$ 
in (\ref{homcount}), we see that inflation models have an intrinsic 
arbitrariness given by the different random magnitudes of $\theta_{\rm i}$ in
different inflationary domains\mycite{Pi84}.  While a large value of $\fa>\!>10^{12}$GeV 
might have been thought to be observationally excluded, it can actually be 
accommodated in domains where $\theta_{\rm i}<\!<1$.  
This may seem highly 
unlikely but, if we consider an infinite inflationary manifold or a 
multiple universe scenario including `all possible worlds', then 
life as we know it would be excluded from those domains with large
$\theta_{\rm i}={\cal O}(1)$ 
because the baryon-to-axion ratio would be too low\mycite{Lin88}.  
Thus, accepting this anthropic selection effect, we have to 
concede that axions could be the dark matter $\Omega_{\rm a} \approx 1$ 
if we live in a domain with
a `tuned' $\theta$-parameter\footnote{This is not quite in the spirit of 
the original motivation for the axion!}:
\be
\theta_{\rm i} ~\approx~ 
0.3\,\Delta^{-1/2} h\,\left(\ma\over10^{-6}\hbox{eV}\right)^{0.6}\,.
\label{anthropic}
\ee
For $\theta_{\rm i} \approx {\cal O}(1)$, this suggests an axion with
$\ma \sim 5\mu$eV ($h=0.5$), though actually inflation makes no 
definite prediction from 
(\ref{anthropic}) beyond apparently specifying $\ma \gapp 10\,\mu$eV. But even 
this restriction is not valid; if we observe (\ref{homcount}) 
carefully 
we see that we can also obtain a dark matter axion for higher $\ma$ by 
fine-tuning $\theta_{\rm i}$ near $\pi$. The anharmonic term ${\rm f}(\theta)$
with an apparent logarithmic divergence allows $\Omega_{\rm a}\approx 1$ 
for a much heavier dark matter axion\mycite{SheBat98}.

This simple `anthropic tuning' picture is significantly altered 
by quantum effects.  
Like any minimally coupled massless 
field during inflation, the axion will have a spectrum of 
quantum excitations associated with
the Gibbons-Hawking temperature $T \sim H/2\pi$.  This implies the field 
will acquire fluctuations about its
mean value $\theta_{\rm i}$ of magnitude
$\delta\theta = H/2\pi\fa$,
giving an effective rms value $\theta_{\rm eff}^2 = (\theta_{\rm i}
+\delta \theta)^2$.  Even if our
universe began in an inflationary domain with $\theta_{\rm i}=0$, there will be
a minimum misalignment angle set by $\delta\theta$; this 
implies that we cannot always fine-tune $\theta_{\rm i}$ in (\ref{anthropic}) 
such that $\Omega_a \lapp 1$.  Worse still, the fluctuations 
$\delta\theta$ imply isocurvature fluctuations in the axion 
density\mycite{isocurv} given by  
(\ref{homcount}), that is, $\delta \rho_{\rm a} 
\propto \theta_{\rm eff}^2
\approx \theta_{\rm i}^2+2\theta_{\rm i}\delta\theta+\delta\theta^2$. Such 
density fluctuations will also create cosmic microwave background (CMB) 
anisotropies which are 
strongly constrained, $\delta T/T \sim \delta \rho/\rho \sim 2\delta\theta
/\theta_{\rm eff}\lapp 10^{-5}$.  Combining the quantum fluctuation $\delta\theta$
with the $\theta$-requirement for a dark matter axion (\ref{anthropic}), 
we obtain a strong constraint on the Hubble parameter during 
inflation\mycite{Lyt90}
\be 
H_1~\lapp ~ 
10^9 \hbox{GeV} \left(10^{-6}\hbox{eV} \over  \ma\right)^{0.4}\,, 
\label{infbound}
\ee
which is valid for small masses $\ma \lapp 10\,\microeV$.
Here, $H_1$ is the Hubble parameter as fluctuations associated with the
time when the 
microwave anisotropies first leave the horizon some 50--60 e-foldings before
the end of inflation.
We conclude from (\ref{infbound}) that inflation in the $\fa>H_1>T_{\rm reh}$
regime  must have a
small Hubble parameter $H_1$ or the dark matter axion must be extremely 
light, $\ma <\!\!< 1\,\microeV$.  Inflation now has many guises and 
models exist with Hubble parameters anywhere in the range 
$H_1 \sim 10^2\hbox{--} 10^{14}$GeV, so a detectable inflationary 
axion with $\ma \sim 1\, 
\microeV$ is possible in principle. Note, however, that the constraint
(\ref{infbound}) can be circumvented in more complicated multifield inflation 
models\mycite{Lin91}.

Quantum effects also constrain the `anthropic tuning' $\theta_{\rm i} 
\rightarrow \pi$ which is required for a larger mass axion $\ma \gapp 10 \,
\microeV$.  For simple inflation models, excessive isocurvature fluctuations 
imply the {\it dark matter} axion mass is bounded above by $\ma \lapp 
1\,$meV for $H_1\gapp 10^2$GeV\mycite{SheBat98}.  Note that this is a
considerably heavier dark matter axion than the oft-quoted inflationary 
limit $\ma\lapp 5\,\microeV$.

\subsection{Axion string network creation during inflation \it (Case 2: $H_1\gapp \fa$)}
Even for a low reheat temperature $T_{\rm reh}< \fa$, one can envisage
inflation models with a large Hubble parameter during inflation
$H_1\gapp \fa$ (such as chaotic inflation with $H_1 \sim 10^{13}-10^{14}$GeV).  
The quantum fluctuations in this case
are sufficient to take the Peccei-Quinn field over the top of the 
potential leaving large spatial variations and topologically 
non-trivial windings 
\mycite{LytSte92}.  We can interpret this 
as the Gibbons-Hawking temperature `restoring' the PQ-symmetry 
$T_{\rm GH} \gapp \fa$.  
As inflation draws to a close and $H$ falls below $\fa$, these 
fluctuations will become negligible and a string network will form.
Provided inflation does not continue beyond this point for more than 
about another 30 e-foldings, we will effectively return to the `standard
thermal' scenario in which axions are produced by a decaying string 
network.  So such low reheat 
inflation models are again only compatible with a dark matter axion, $\ma 
\approx 100\mu$eV. 

We note that there is also a borderline
scenario with $H_1\sim \fa$ and in which domain walls form but few strings.
Since strings are required to remove them, such domian walls will be  
cosmologically unacceptable\mycite{LinLyt90}.

\section{CONCLUSIONS}

\noindent We have endeavoured to provide an overview of axion cosmology 
focussing on the mass of a dark matter axion.  First, 
the cosmological axion density was considered
 in the standard thermal scenario where the dominant
contribution comes from axion strings.  In this case there is, in principle, 
a well-defined calculational
method to precisely predict the mass $\ma$ of a dark matter axion.
For the currently favoured value of the Hubble parameter ($H_0 \approx 
60\rm  \,km\,s^{-1}Mpc^{-1}$), the estimate 
(\ref{stringbound}) predicts a dark matter axion of mass $
\ma \approx 200\,\mu\hbox{eV}$, where 
significant uncertainties from all sources approach an order of 
magnitude.  
The additional uncertainty in this string calculation is the parameter ratio
$\alpha/\kappa\approx1$, that is, the ratio of the loop size to radiation backreaction 
scale.

Secondly, we have reviewed inflationary axion cosmology showing that 
(i) many inflation models return us to the standard thermal scenario 
with $\ma \sim 100\mu$eV, (ii) some inflation models are essentially 
incompatible with a detectable dark matter axion and (iii),  
because of the possibility of 
`anthropic fine-tuning', other inflation models
can be constructed which 
incorporate a dark matter axion mass 
anywhere below $\ma\lapp 1$meV. 

We conclude that, while a dark matter axion might possibly lurk anywhere
in an enormous mass range below $\ma \lapp 1\,$meV, the best-motivated mass 
for future axion 
searches lies near $\ma \sim 100\mu$eV, a standard thermal scenario 
prediction which is also 
compatible with a broad class of inflationary models.  

\section*{ACKNOWLEDGEMENTS}

We are grateful to Pierre Sikivie for his kind invitation to attend
the Florida Axions Workshop, at which we benefitted from 
many informative discussions
with members of the theoretical and experimental axion community. 
We acknowledge useful conversations with Andrei Linde,
David Lyth and Georg Raffelt. 
This work is supported by PPARC.


\def\jnl#1#2#3#4#5#6{\hang{#1, {\it #4\/} {\bf #5}, #6 (#2).}}


\def\jnlerr#1#2#3#4#5#6#7#8{\hang{#1, {\it #4\/} {\bf #5}, #6 (#2).
{Erratum:} {\it #4\/} {\bf #7}, #8.}}


\def\jnltwo#1#2#3#4#5#6#7#8#9{\hang{#1, {\it #4\/} {\bf #5}, #6 (#2);
{\it #7\/} {\bf #8}, #9.}}

\def\prep#1#2#3#4{\hang{#1 (#2),  #4.}}

\def\myprep#1#2#3#4{\hang{#1 (#2), '#3', #4.}}

\def\proc#1#2#3#4#5#6{\hang{#1 (#2), `#3', in {\it #4\/}, #5, eds.\ (#6).}
}
\def\procu#1#2#3#4#5#6{\hang{#1 (#2), in {\it #4\/}, #5, ed.\ (#6).}
}

\def\book#1#2#3#4{\hang{#1 (#2), {\it #3\/} (#4).}
									}

\def\genref#1#2#3{\hang{#1 (#2), #3}
									}


\def\prl{Phys.\ Rev.\ Lett.}
\def\pr{Phys.\ Rev.}
\def\pl{Phys.\ Lett.}
\def\np{Nucl.\ Phys.}
\def\prp{Phys.\ Rep.}
\def\rmp{Rev.\ Mod.\ Phys.}
\def\cmp{Comm.\ Math.\ Phys.}
\def\mpl{Mod.\ Phys.\ Lett.}
\def\apj{Ap.\ J.}
\def\apjl{Ap.\ J.\ Lett.}
\def\aap{Astron.\ Ap.}
\def\cqg{Class.\ Quant.\ Grav.} 
\def\grg{Gen.\ Rel.\ Grav.}
\def\mn{M.$\,$N.$\,$R.$\,$A.$\,$S.}
\def\ptp{Prog.\ Theor.\ Phys.}
\def\jetp{Sov.\ Phys.\ JETP}
\def\jetpl{JETP Lett.}
\def\jmp{J.\ Math.\ Phys.}
\def\cupress{Cambridge University Press}
\def\pup{Princeton University Press}
\def\wss{World Scientific, Singapore}


\begin{thebibliography}{9}

\bibitem{PecQui77}
\jnltwo{R.D. Peccei and H.R. Quinn}{1977}{CP conservation and the 
presence of pseudoparticles}{\prl}{38}{1440}{\pr}{D16}{1791}
\jnl{F. Wilczek,}{1978}{Problem of strong $P$ and $T$ invariance in the presence of instantons}{\prl}{40}{279}
\jnl{S. Weinberg}{1978}{A new light boson?}{\prl}{40}{223}



\bibitem{Raf97}
\prep{G.G. Raffelt}{1997}{}{astro-ph/9707268}

\bibitem{Hagetal98}
\prep{C. Hagmann {\it et al}}{1998}{}{astro-ph/9802061}
\jnl{I. Ogawa, S. Matsuki \& K. Yamamoto}{1996}{}{\pr}{D53}{R1740}
Refer to other papers in these proceedings.

\bibitem{VilEve82}
\jnl{A. Vilenkin and A.E. Everrett}{1982}{}{\prl}{48}{1867}

\bibitem{Dav86}
\jnl{R.L. Davis}{1986}{}{\pl}{180B}{225}
\jnl{R.L. Davis and E.P.S. Shellard}{1989}{Do axions need inflation?}
{\np}{B324}{167}

\bibitem{Sik82}
\jnl{P. Sikivie}{1982}{Axions, domain walls and the early universe}{\prl}{48}{1156}


\bibitem{She86}
\procu{E.P.S. Shellard}{1986}{Axionic domain walls \& cosmology}{{\rm Proceedings
of the 26th Liege International Astrophysical Colloquium,}}{Demaret, J.}
{University de Liege} \proc{E.P.S. Shellard}{1990}{Axion strings and domain walls}{Formation and
Evolution of Cosmic Strings}{Gibbons, G.W., Hawking, S.W., \& Vachaspati,
V.}{\cupress}


\bibitem{PAD83}
\jnl{J. Preskill, M.B. Wise and F. Wilczek}{1983}{Cosmology of the invisible
axion}{\pl}{120B}{127}
\jnl{L. Abbott \& P. Sikivie}{1983}{}{\pl}{120B}{133}
\jnl{M. Dine \& W. Fischler}{1983}{}{\pl}{120B}{137}

\bibitem{Tur86}
\jnl{M.S. Turner}{1986}{Cosmic and Local Mass Density of ``Invisible'' 
Axions}{\pr}
{D33}{889}

\bibitem{SheBat98}
\genref{E.P.S Shellard and R.A. Battye}{1998}{Cosmic Axions}{astro-ph/9802216}
\prep{E.P.S. Shellard \& R.A. Battye}{1998}{}{`Inflationary axion cosmology 
revisited', DAMTP preprint}

\bibitem{BatShe94a}
\jnl{R.L. Davis and E.P.S. Shellard}{1988}{Antisymmetric tensors and
spontaneous symmetry breaking}{\pl}{214B}{219}
\jnl{R.A. Battye and E.P.S. Shellard}{1994}{Global
string radiation}{\np}{B423}{260}

\bibitem{VilShe94}
\prep{A. Vilenkin \& E.P.S.\ Shellard}{1994}{}{{\it Cosmic strings and other 
Topological Defects} (\cupress)}

\bibitem{BBAS90}
\jnl{D.P. Bennett and F.R. Bouchet}{1990}{High resolution simulations of cosmic
string evolution: network evolution}{\pr}{D41}{2408}
\jnl{B. Allen and E.P.S. Shellard}{1990}{}{\prl}{64}{119}

\bibitem{BatShe94b}
\jnlerr{R.A. Battye and E.P.S Shellard}{1994}{}{\prl}{73}{2954}{76}{2203}

\bibitem{BatShe98b}
\genref{R.A. Battye and E.P.S Shellard}{1998}{`Spectrum of radiation from 
axion strings' in these proceedings.}

\bibitem{Lyt92}
\jnl{D.H. Lyth}{1992}{Estimates of the cosmological axion density}{\pl}
{275B}{279}

\bibitem{Nag97}
\jnl{M. Nagasawa}{1997}{}{Prog.\ Theor.\ Phys.}{98}{851}
\jnl{M. Nagasawa \& M. Kawasaki}{1994}{}{\pr}{D50}{4821}

\bibitem{Pi84}
\jnl{S.Y-.~Pi}{1984}{}{\prl}{52}{1725}

\bibitem{Lin88}
\jnl{A.D. Linde}{1988}{Inflation and axion cosmology}{\pl}{201B}{437}

\bibitem{isocurv}
\jnl{M.S. Turner, F. Wilczek \& A. Zee}{1983}{}{\pl}{120B}{127}

\bibitem{Lyt90} 
\jnl{D.H. Lyth}{1990}{}{\pl}{B236}{408}
\jnl{M.S. Turner \& F. Wilczek}{1991}{}{\prl}{66}{5}

\bibitem{Lin91}
\jnl{A.D. Linde}{1991}{Axions in inflationary cosmology}{\pl}{259B}{38}

\bibitem{LytSte92}
\jnl{D.H. Lyth and E.D. Stewart}{1992}{}{\pl}{283B}{189}

\bibitem{LinLyt90}
\jnl{A.D. Linde \& D.H. Lyth}{1990}{}{\pl}{246B}{353}





\end{thebibliography}
\end{document}